\documentclass[prl,aps,twocolumn,superscriptaddress,nofootinbib]{revtex4-2}
\pdfoutput=1
\usepackage{amsmath, amsthm, amssymb}
\usepackage{amsfonts}
\usepackage{graphicx}
\usepackage{dcolumn}
\usepackage{bm}
\usepackage{textcomp}
\usepackage[normalem]{ulem}

\usepackage[titletoc,title]{appendix}

\usepackage{epsfig}
\usepackage{subfigure}
\usepackage{url}
\usepackage{float}
\usepackage{cases}

\newcommand{\dd}{\mathrm{d}}

\usepackage{colordvi}
\usepackage[usenames,dvipsnames]{xcolor}

\usepackage[colorlinks=true, urlcolor=blue, anchorcolor=blue, citecolor=blue,filecolor=blue,linkcolor=blue,menucolor=blue
]{hyperref}

\usepackage[titletoc,title]{appendix}

\usepackage{color}
\long\def\magenta#1{#1}

\begin{document}
\title{Emergent population dynamics of random walkers with cooperative reproduction and spatial selection}

\author{Ohad Vilk}
\email{ohad.vilk@mail.huji.ac.il}
\affiliation{Racah Institute of Physics, Hebrew University of
Jerusalem, Jerusalem 91904, Israel}

\author{Baruch Meerson}
\email{meerson@mail.huji.ac.il}
\affiliation{Racah Institute of Physics, Hebrew University of
Jerusalem, Jerusalem 91904, Israel}

\begin{abstract}

We extend the $N$ branching Brownian motions model of population invasion to higher-order asexual reproduction. Increasing reproduction order leads to  qualitative changes: invasion fronts generically cease to exist beyond binary reproduction; and in the binary case itself, their speed becomes diffusion-independent. Ternary reproduction shows critical behavior, with collapse into a strongly localized `invasion bullet' in the supercritical regime, diffusive spreading in the subcritical regime, and a continuous family of fronts at criticality. These results suggest that the dominance of division and binary reproduction in nature reflects fundamental constraints on invasion dynamics. 
 
\end{abstract}
\maketitle

Invasion of a population into an unoccupied habitat is a fundamental process in ecology and evolution \cite{MacArthur,Kot}, and its modeling has attracted considerable attention in physics and mathematics starting from the classical papers  of Fisher \cite{Fisher} and Kolmogorov, Petrovskii and Piscounov (KPP) \cite{KPP}. A minimal 
model of invasion by a fixed-size population is the 
$N$-BBM ($N$ branching Brownian motions)  
model with spatial selection \cite{Maillard,DeMasi,BBD}. It describes $N\gg 1$ Brownian particles on the line, each of which can branch into two particles: $A\to 2A$.  At every branching the leftmost  particle is eliminated, keeping the
population size constant. In the hydrodynamic (HD) limit $N\!\to\! \infty$, the dynamics of the macroscopic population density $u(x,t)$, which we normalize to 1, is described by a moving-boundary problem \cite{DeMasi,BBD}:
\begin{eqnarray}
&&\partial_t u = \lambda u+D \partial_x^2 u, \quad x>L(t), \label{pde0}\\
&&u[x=L(t),t] = 0,\quad \int_{L(t)}^{\infty} u(x,t) dx=1\,,\label{conservation0}
\end{eqnarray}
where $D$ is the diffusion constant and $\lambda$ is the branching rate. The HD formulation is completed by specifying an initial condition $u(x, t=0)$.  Spatial selection enters as a moving absorbing boundary at $x=L(t)$, determined implicitly by particle conservation. The $N$-BBM model also describes adaptation of a population undergoing mutations along a fitness axis,  with selection removing the least fit individual \cite{BD1997}. 

At long times this population 
develops an invasion wave: a traveling wave solution (TWS) propagating with the speed $c=2\sqrt{\lambda D}$. It propagates into a linearly unstable state $u=0$ and belongs to the universality class of \emph{pulled} waves \cite{Saarloos2003,Scheel}, the best known member of which is the Fisher-KPP wave \cite{Fisher,KPP}.   When $N$ is finite,  typical fluctuations of pulled waves  scale logarithmically with $N$ and are therefore very large. They originate in the wave's leading edge and are dominated by a few particles -- the front runners \cite{BD1997,Pechenik,BD19992001,Panja2003,Derridaetal2006}.  Typical fluctuations of pulled waves have attracted much interest, see Refs. \cite{Saarloos2003,Panja,Kuehn} for reviews. Large deviations of the empirical speed of pulled waves at long times have also been studied \cite{MS2011,MVS2013,MS2024}. 

\begin{figure}[t]
\vspace{-.2cm}
\includegraphics[width=\columnwidth,clip=]{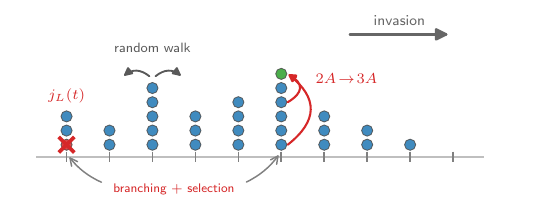}
\vspace{-.8cm}
\caption{\label{fig:illustration} RW with binary reproduction and spatial selection.}
\vspace{-.2cm}
\end{figure}

Returning to the biological motivation behind the $N$-BBM model, we observe that the branching  process $A\to 2A$ describes the simplest way of reproduction: by division. How would the invasion dynamics change in the presence of asexual \emph{cooperative} reproduction, 
e.g. binary $2A\to 3A$ or ternary $3A \to 4A$ branching processes? A natural setting here involves $N\gg 1$ continuous-time random walkers (RWs) on a regular one-dimensional lattice $j=\dots,-1,0,1, \dots$. Any $k$ RWs on the same lattice site can take part in the branching reaction $kA \to (k+1)A$. When a new particle is born, one of the particles at the leftmost populated lattice site $j=j_L(t)$ is removed. The model is illustrated in Fig.~\ref{fig:illustration} for $k=2$.

Here we show that higher-order reproduction leads to important 
consequences for the invasion dynamics. 
For the binary branching $2A \to 3A$ the RWs do form a robust macroscopic TWS, albeit a TWS quite different from its counterpart for $A\to 2A$. First,  for large $D$ the wave propagation speed becomes $D$-independent,  and is determined solely by the branching rate. Second, this wave is \emph{strongly pushed}:  it propagates into a state which is linearly stable, but nonlinearly unstable with a zero instability threshold. The wave properties in this regime are determined by all the particles rather than by a few front runners. As a result, such a wave exhibits the customary $1/N$ scaling of the wave speed shift and fluctuations \cite{HZ2026}.

We find that, generically, for branching of higher order than binary, no TWSs exist. In particular, for ternary branching $3A\to 4A$ the existence of macroscopic TWSs requires fine tuning of the ratio of the branching rate and the diffusion constant. In the absence of fine tuning the population either spreads diffusively, 
or collapses into an `invasion bullet' whose size is comparable with the lattice spacing. 
For quaternary, $4A\to 5A$, and higher-order branchings, the population does not exhibit any TWSs. It ultimately spreads by diffusion up to subleading corrections coming from the reproduction.

\emph{HD model.} When the number of particles on a lattice site $j$ is large, $n_j\gg 1$, we can approximate the combinatorial rate of the $kA \to (k+1)A$ branching by its leading-order term $\Lambda n_j^k/N^{k-1}$ \cite{Gardiner}, where  $\Lambda=\lambda/k!$, and the factor $N^{k-1}$ is introduced  for convenience.  Of  most interest is the macroscopic regime, where the characteristic spatial length scale is much larger than the lattice 
spacing $h$. In this regime we can approximate the RW by continuous diffusion and arrive at the following HD model for the coarse-grained population density $u(x,t)$:
\begin{eqnarray}
&&\partial_t u = \Lambda u^k+D \partial_x^2 u, \quad x>L(t), \label{pde1}
\end{eqnarray}
alongside Eq. \eqref{conservation0}. Here $D = D_0h^2$ is the diffusion coefficient, and $D_0$ is the hopping rate. 
As in the $N$-BBM model (\ref{pde0}),  the position $x=L(t)$ of the moving absorbing boundary
is determined implicitly by mass conservation. The HD model (\ref{pde1}) generalizes the $N$-BBM model to arbitrary integer $k$.  

\emph{Dimensional analysis.} Like in many other problems \cite{Barenblatt},  a valuable insight is provided by a simple dimensional analysis. Indeed, the only dimensional parameters entering Eqs.~(\ref{conservation0}) and~(\ref{pde1}) 
are $\Lambda$,  with units $\text{length}^{k-1}/\text{time}$, and $D$, with units $\text{length}^2/ \text{time}$. For $k\neq 3$ these parameters define length  and time scales, 
\begin{equation}
\label{lengthtime}
\ell=\left(D/\Lambda\right)^{\frac{1}{3-k}} \quad \text{and} \quad  \tau=\left(D^{k-1}/\Lambda^2\right)^{\frac{1}{3-k}}\,.
\end{equation}
Therefore, if there are traveling waves in this model, their speed $c$ must scale as 
\begin{equation}
\label{speed0}
c\sim  \ell/\tau= \left(\Lambda D^{2-k}\right)^{\frac{1}{3-k}}\,.
\end{equation}
For $k=1$ this yields $c=a\sqrt{\Lambda D}$, in agreement with the exact result for the $N$-BBM model, where $a=2$.   

Remarkably, for $k=2$ Eq.~(\ref{speed0}) yields a counterintuitive prediction $c\sim \Lambda$, independent of $D$.  We will verify this prediction and compute the numerical prefactor. 

The special case  $k=3$ is dimensionally deficient. Here $\Lambda$ and $D$ have the same units of $\text{length}^2/\text{time}$, so Eqs.~\eqref{conservation0} and~(\ref{pde1})  do not define any intrinsic length or time scale. As we show below, the long-time dynamics of this system is controlled by the dimensionless parameter $\alpha=\Lambda/D$. 

For $k>3$ the dimensional analysis leading to Eq.~(\ref{lengthtime})  is not very useful because, at long times, the reproduction term in Eq.~(\ref{pde1}) -- and the parameter $\Lambda$ -- become irrelevant. The ensuing long-time population spread is described, up to small corrections, by the simple diffusion equation.  Now we consider the cases of $k=2,3, \dots$, separately. 

\begin{figure}[t]
\includegraphics[width=0.75\columnwidth,clip=]{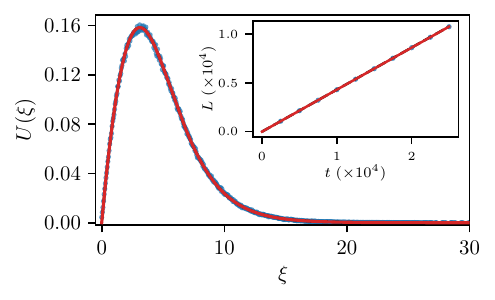}
\vspace{-.4cm}
\caption{\label{fig:k2} Binary reproduction ($k\!=\!2$): rescaled TWS $U(\xi)$. Red solid line: numerical solution of Eq.~(\ref{ode}), obeying Eq.~(\ref{conservation0}). Black dashed line: numerical solution of Eqs.~(\ref{pde2}) and (\ref{conservation0}) at long times. Blue circles: MC simulation for $D\!=\!10^5$ and $N\!=\!10^4$ at $t\!\simeq\! 2.25\!\times\!10^4$. Inset: front position $L(t)$ vs. time. All three methods yield $c\!\simeq\! 0.43$.}
\end{figure}

\emph{Binary reproduction, $k=2$}.  In this case Eq.~(\ref{lengthtime}) yields $\ell= D/\Lambda$ and $\tau = D/\Lambda^2$. The rescaling transformation $x'=x/\ell$, $t'=t/\tau$ and $u'=\ell u$ brings \eqref{pde1} into a parameter-free dimensionless form
\begin{eqnarray}
&&\partial_t u = u^2+\partial_x^2 u , \quad x>L(t).\label{pde2}
\end{eqnarray}
Here, and in Eq. \eqref{conservation0}, $L(t)$ is rescaled by $\ell$, and we dropped the primes. 
At long times, $t\!\gg \!1$, the solution to the rescaled problem (\ref{conservation0}) and (\ref{pde2})  approaches a unique TWS $u(x,t\!\gg\! 1) = U(x\!-\!ct) \equiv U(\xi)$, and $L(t) = ct+\text{const}$. This solution is described by the  ordinary differential equation (ODE)
\begin{equation}
U^{\prime\prime} (\xi) + c U^{\prime}(\xi)+U^2=0, \quad \xi>0,\label{ode}
\end{equation}
with normalization condition $\int_{0}^{\infty} U(\xi) d\xi =1$.
For any $c>0$, Eqs.~(\ref{ode}) with $U(0)=0$ can be solved numerically by demanding the asymptotic behavior
$U(\xi\!\to\!\infty)\! \sim\! e^{-c\xi}$ at infinity, where the $U^2$ term becomes negligible. The speed $c$ can then be determined iteratively so that the normalization condition is obeyed to a desired precision. We find $c\simeq0.43$, and the resulting density profile $U(\xi)$ is shown in Fig.~\ref{fig:k2} alongside  the late-time density profiles obtained by (a) solving the full HD problem \cite{SM} and (b) a Monte-Carlo (MC) simulation of the lattice model \cite{SM}. 

Back in the original variables, the traveling wave speed $c\simeq 0.43 \Lambda$ is independent of the diffusion coefficient $D$, as predicted from the dimensional analysis. Note, however, that 
this result holds only when the characteristic spatial extension of the wave $\ell$ is much larger than the lattice spacing $h$, ensuring the validity of the continuum model.  Since $\ell=D/\Lambda$, this condition reads $D\gg \Lambda h$ or, in terms of the original microscopic model, $D_0 h\gg \Lambda$.

\begin{figure*}[t]
\includegraphics[width=0.9\textwidth,clip=]{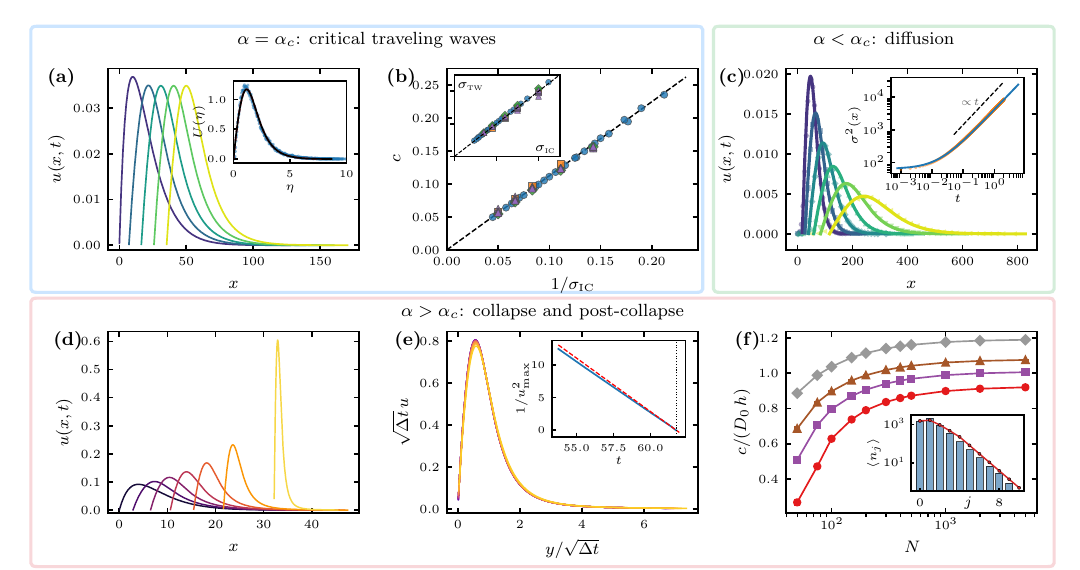}
\vspace{-.2cm}
\caption{\label{fig:k3} Ternary reproduction ($k=3$).
Solid lines: HD numerics~\cite{SM}; symbols: MC simulations, unless noted otherwise.
(a,\,b)~$\alpha=\alpha_c$: TWSs.
(a)~$u(x,t)$ vs.\ $x$ at five equally spaced times from $t=0$ to $t\simeq 475$ ($\alpha_c\simeq 7.089$, $D=1$). Inset: rescaled TWS $U(\xi)$, see Eq.~(\ref{ODE3a}) (black), the HD solution at late times [Eq.~(\ref{family3}), red dashed], and MC (blue circles).
(b)~TWS speed selection: $c$ vs.\ $1/\sigma_{\text{IC}}$ from HD runs with gamma, Gaussian, half-sine, and stretched-exponential initial conditions. Dashed line: linear fit $c\propto 1/\sigma_{\text{IC}}$. Inset: $\sigma_{\rm TWS}$ vs.\ $\sigma_{\rm IC}$ (see text).
(c)~$\alpha<\alpha_c$  ($\alpha=5$): diffusive spreading. Shown is $u(x,t)$ vs.\ $x$ at six log-spaced times from $t\simeq 0.1$ to $t\simeq 2$ (MC: $N=500$). Inset: $\sigma^2[u]$ vs.\ $t$ (log-log); dashed: $\propto\! t$.
(d--f)~$\alpha>\alpha_c$: collapse and post-collapse.
(d)~$u(x,t)$ vs.\ $x$ at $t=0, 10, 20, \ldots, 60$ ($\alpha=9$, $D=1$).
(e)~Self-similar collapse: $\sqrt{\Delta t}\,u$ vs.\ $y/\sqrt{\Delta t}$ [Eq.~(\ref{similarity1})] at twelve times near $t_c\simeq 61.8$, with $\Delta t$ ranging from $0.3$ to $3$. Inset: $1/u_{\max}^2$ vs.\ $t$, showing linear behavior near $t_c$.
(f)~Post-collapse discrete TW: rescaled speed $c/(D_0\,h)$ vs.\ $N$ (MC; $\alpha=8$, $8.5$, $9$, $10$). Inset: mean densities $\langle n_j\rangle$  from MC (bars) are compared with discrete mean-field theory (red lined circles) for $N=5000$, $\alpha=9$ ($j$ is measured relative to $j_L$).}

\end{figure*}

\emph{Ternary reproduction, $k=3$.} In this case there is \emph{one} special value of the parameter  $\alpha\equiv\Lambda/D$ for which the HD equations~\eqref{conservation0} and (\ref{pde1})  have a continuous family of TWSs with arbitrary speed $c>0$. To show this, we make the traveling wave ansatz $u(x,t) = u(x-ct)$, which transforms Eq. \eqref{pde1} into the ODE
\begin{equation}\label{ODE3}
 D u^{\prime\prime}(\xi) + c u^{\prime}(\xi)  +\Lambda u^3=0\,,\quad \xi>0\,.
\end{equation}
In the presence of a new dimensional parameter $c$ we can rescale the variables: $\xi'=\xi/\ell_0$, and $U=\beta u$, where $\ell_0 =D/c$ and $\beta=  \sqrt{\Lambda D}/c$. As a result,  Eq.~(\ref{ODE3}) becomes dimensionless and parameter-free:
\begin{equation}\label{ODE3a}
U^{\prime\prime}(\xi)+U^{\prime}(\xi)  +U^3=0\,,\quad \xi>0\,,
\end{equation}
where we have omitted the primes. The rescaled mass conservation condition reads
\begin{equation}\label{mass3}
\int_0^{\infty} d\xi\,U(\xi) =\sqrt{\alpha}\,, \quad \text{where}\quad \alpha=\Lambda/D\,.
\end{equation}
As one can see, the traveling wave speed $c$ drops from the rescaled formulation. 
We solved Eq.~(\ref{ODE3a}) numerically  with the boundary conditions $U(0)=0$  and $U(\infty) =0$.  Then, evaluating the integral (\ref{mass3}), we obtained $\alpha =\alpha_c \simeq 7.089$: the only value of the parameter $\alpha$  for which these TWSs exist.  The rescaled TWS $U(\xi)$ is shown in the inset of Fig.~\ref{fig:k3}(a).  Back in the original variables the continuous family of TWSs for $k=3$ can be written as
\begin{equation}
    \label{family3}
    u(x,t;c) =\frac{c}{\alpha_c D} U\left[\frac{c}{D}\left(x-ct\right)\right]\,,\quad x\geq ct\,,
\end{equation}
with arbitrary $c>0$. 

In the full HD problem (\ref{conservation0})-(\ref{pde1}) -- which can be solved numerically \cite{SM} -- a particular TWS is selected by the initial condition, as illustrated in Fig.~\ref{fig:k3}(a).  Very similar results are obtained in the MC simulations. What feature of the initial condition selects a particular TWS is an interesting open question
which deserves a separate study. We observed, however, that the characteristic width $\sigma_{\text{IC}}\equiv \sigma[u(x,0)]$ of the initial condition $u(x,t=0)$ provides a surprisingly good empirical predictor of a selected TWS.  Here we defined $\sigma$ by analogy with the variance of a probability distribution
$\sigma^2[u(x,t)] \equiv \int_{L(t)}^{\infty}x^2 u dx-(\int_{L(t)}^{\infty}x u dx)^2.$
For a TWS this becomes, in the rescaled form,
$\sigma^2_{\text{TWS}}\equiv \int_0^{\infty}\xi^2 U d\xi-(\int_0^{\infty}\xi U d\xi)^2.$
Fig.~\ref{fig:k3}(b) shows that the selected speed $c$ approximately scales as $c\propto 1/\sigma_{\text{IC}}$, while the inset of  this panel
confirms that the characteristic width of the TWS $\sigma_{\text{TWS}}$ remains fairly close to the initial width $\sigma_{\text{IC}}$ as time progresses. We have checked this for different families of initial conditions, see the caption of Fig.~\ref{fig:k3}(b).

For $\alpha \neq \alpha_c$ the ternary reproduction cannot support macroscopic TWSs. Numerical solutions of Eqs.~(\ref{conservation0}) and~(\ref{pde1}) demonstrate two distinct regimes: $\alpha < \alpha_c$ and $\alpha > \alpha_c$. For $\alpha < \alpha_c$, the population spreads diffusively, but the reproduction is still relevant, see \cite{SM} for details. Figure~\ref{fig:k3}(c) illustrates this sub-critical diffusive regime for 
$\alpha=5$. Panel (c) also compares the HD solution with MC simulations, showing a good agreement, while the inset confirms, for both the HD model and the microscopic one,  the late-time diffusion scaling $\sigma^2[u]\propto t$.

For $\alpha>\alpha_c$, the density $u(x,t)$, as described by the HD model, exhibits a finite-time collapse to a point.  To take a closer look into the collapse regime, we pass to the reference frame moving with the absorbing boundary, $y=x-L(t)$.  In this frame Eq.~(\ref{pde1}) reads
\begin{equation}\label{pdequat3a}
\partial_t u =\Lambda u^3+D \partial_y^2 u +\dot{L}\partial_y u, \quad y>0\,.
\end{equation}
Close to the collapse time $t_c$ (which depends on the initial condition), and not too far from the moving boundary $L(t)$ (see below),
the solution of Eq.~(\ref{pdequat3a}) becomes self-similar. Indeed,  by applying dimensional analysis \cite{Barenblatt} and assuming that the scaling behavior is unaffected by the length scale introduced by the initial condition $u(x,t=0)$, we arrive at the following similarity ansatz:
\begin{equation}
\label{similarity1}
u(y,t) \!=\! \frac{1}{\sqrt{D\Delta t}} V\left(\frac{y}{\sqrt{D\Delta t}}, \alpha\right),\;
\dot{L}\!= \!a(\alpha)\sqrt{\frac{D} {\Delta t}},
\end{equation}
where $\Delta t\equiv t_c-t$ is the remaining time until the singularity, and $\alpha=\Lambda/D$ as before. It is crucial that this ansatz is compatible with the mass conservation law (\ref{conservation0}). 
The coordinate of the moving boundary, close to the singularity, is 
$L(t)=L_c -2a(\alpha) \sqrt{D\Delta t}$, where $L_c$ depends on the initial condition.   Figure~\ref{fig:k3}(d) shows $u(x,t)$ vs. $x$ at different times for $\alpha=9$, obtained numerically. Panel (e) verifies the similarity ansatz quantitatively. As one can see, the density profiles near $t_c$ collapse onto a single curve when plotted in the self-similar variables, while the inset shows the expected linear dependence $1/u_{\max}^2\propto\Delta t$.  

The scaling function $V(z)$ obeys the nonlinear ODE 
\begin{equation}
\label{simODE}
2 V''(z)+(2 a-z) V'(z)+2 \alpha V(z)^3-V(z) \!=\!0,\; z>0,
\end{equation}
with the boundary condition $V(0)\!=\!0$, while the constant $a=a(\alpha)$ is a ``nonlinear eigenvalue" of this problem \cite{Barenblatt}. 

Notably, the self-similar solution (\ref{similarity1})  does not provide a full solution for $u(x,t)$ near the collapse, as it cannot satisfy the mass conservation condition $\int_0^{\infty} V(z)\, dz=1$.
In order to see why, let us determine the asymptotic behavior of  $V(z)$ at $z\gg 1$. Here we can neglect the nonlinear term $2 \alpha V(z)^3$ in Eq.~(\ref{simODE}) and obtain a linear equation which has the form of a full derivative.
Solving it, we see that the only solutions that decay at $z\to \infty$ behave as 
\begin{equation}
\label{statictail1}
V(z)\simeq C(\alpha)(z-2a)^{-1}.
\end{equation}
The slow decay with $z$ causes a logarithmic divergence of the integral $\int_0^{\infty} V(z) dz$. This implies that, at sufficiently large $z$, the self-similar solution (\ref{similarity1}) needs to be matched to a non-self-similar \emph{outer} solution $v(y,t)$ of Eq.~(\ref{pdequat3a}) which decays sufficiently rapidly  at $y\to \infty$ and keeps the memory of the initial condition $u(x,t=0)$.  Please see Ref. \cite{SM} for a ``road map" to this matching.  Importantly, the dynamical scaling exponents $1/2$ of the self-similar collapse, see  Eq.~(\ref{similarity1}),  are universal -- that is, independent of the initial condition. The coefficients $a(\alpha)$ and $C(\alpha)$ are not universal: they are determined -- by the mass conservation -- through the matching to the outer solution, which depends on the initial condition, but these non-universal dependencies are only logarithmic. 

When the characteristic length scale $\sim (D\Delta t)^{1/2}$ of the collapsing population becomes comparable with the lattice spacing, the HD theory breaks down.  MC simulations reveal the population concentrating into a microscopic `invasion bullet' that advances at a constant speed. 
Figure~\ref{fig:k3}(f) shows the measured rescaled speed $c/(D_0\,h)$ vs $N$ of the invasion bullet for several values of $\alpha>\alpha_c$.  The inset displays the average lattice density profile $\langle n_j\rangle$ for $N=5000$ and $\alpha=9$, when the invasion bullet occupies about $8$ lattice sites, compared to numerical solutions of a discrete mean-field theory~\cite{SM}.

\begin{figure}[t]
\includegraphics[width=0.8\columnwidth,clip=]{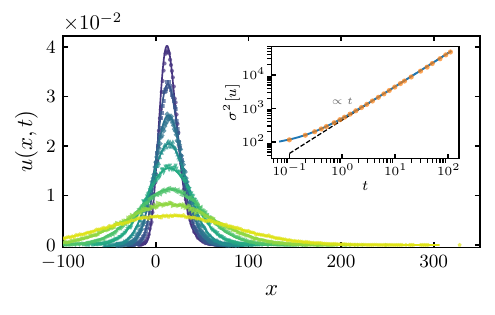}
\vspace{-.4cm}
\caption{\label{fig:k4}Quaternary reproduction ($k=4$): diffusive spreading in HD numerics (solid lines) and MC simulations (symbols, $N=30$, $5000$ replicates). $u(x,t)$ for  $D=225$ ($\ell/h=5$) at eight log-spaced times from $t=0.1$ to $t=10$. Inset:  $\sigma^2[u]$ vs.  $t$ approaching the diffusive scaling $\propto t$ 
(dashed line).}
\end{figure}

\emph{$k>3$: diffusive spread.}
Numerical solutions of the HD equation~(\ref{pde1}) with (\ref{conservation0}), as well as MC simulations
show that for $k>3$  the population spreads by diffusion at long times, up to subleading corrections coming from reproduction. While for $k=3$ the reproduction remains quantitatively relevant, for $k>3$ it only produces  logarithmic corrections to pure diffusion. In particular, for $k=4$ the center of mass drifts to the right as $M(t)\sim (\ln t)^{3/2}$~\cite{SM}.

Figure~\ref{fig:k4} shows the density profiles of $u(x,t)$ at different times for $N=30$ and $D=225$.  At  early times the profile is sharply peaked due to the quaternary reproduction; it then rapidly broadens and acquires a Gaussian shape, characteristic of diffusion. The MC simulations agree quantitatively with the HD results.  The inset confirms that the spatial spread $\sigma^2[u]$ exhibits a linear growth with time, consistent with pure diffusion. Similar results were obtained for $k=5$. 

That the diffusion dominates over the reproduction for $k>3$ follows
from a comparison of the reproduction and diffusion terms in Eq.~(\ref{pde1}).  If the process is diffusion-dominated, the length scale grows as $\sqrt{Dt}$, while the density goes down as $u_{\text{max}}\sim 1/\sqrt{Dt}$. As a result, the reproduction term behaves as $\Lambda u_{\text{max}}^k \sim \Lambda (Dt)^{-k/2}$, so that the ratio 
$\Lambda u^k/(D\partial_x^2 u)$ scales with time as $t^{(3-k)/2}$ and goes to zero for $k>3$.

\emph{Discussion}.  We showed that higher-order reproduction leads to qualitative changes in invasion dynamics. 
In particular, macroscopic TWSs cease to exist for $k>3$, and are generically absent already in the ternary case. The ternary case presents us with interesting questions that merit further study: (i) selection of a unique TWS from a continuous family in the critical case $\alpha=\alpha_c$, and (ii) a full description of the collapse of a supercritical population ($\alpha>\alpha_c$). 

To conclude, highly-simplified models such as the $N$-BBM model and our model for general $k$ do not attempt to explain the empirical fact that reproduction by division and binary
reproduction dominate living systems, whereas higher-order mechanisms are rare \cite{book}. Still, it is fascinating that minimal models like these already encode this preference. Indeed, as we have shown, the ability to sustain robust invasion fronts is a generic property of low-order reproductions, 
but not of higher-order ones.

The research of B.M. is supported by the Israel Science Foundation (Grant No. 1579/25).

\onecolumngrid

\clearpage


\renewcommand{\thesection}{S\arabic{section}}
\renewcommand{\thefigure}{S\arabic{figure}}
\renewcommand{\thetable}{S\arabic{table}}
\renewcommand{\theequation}{S\arabic{equation}}

\setcounter{figure}{0}
\setcounter{table}{0}
\setcounter{equation}{0}

\section*{Supplemental Material for ``Emergent population dynamics of random walkers
  with cooperative reproduction and spatial selection''}

Here we provide some details to support the derivations of the main text. In what follows, the notations and abbreviations are the same as in the main text, and the numbered equations and figures refer to those therein.

\tableofcontents

\clearpage

\section{Hydrodynamic solver}\label{sec:pde}

The hydrodynamic (HD) equations~(2) and~(3) of the main text are solved in the co-moving frame
$y = x - L(t)$, where $L(t)$ is the position of the absorbing
boundary.  Setting $D=1$, the transformed
problem reads
\begin{gather}
  \partial_{\magenta{t}} u = \partial_y^2 u + \dot{L}\,\partial_y u
    + \frac{\Lambda}{D}\,u^k\,,\quad y > 0\,,
    \label{eq:pde_comov}\\
  u(0,\magenta{t}) = 0\,,
    \label{eq:abs_bc}\\
  \partial_y u(0,\magenta{t}) = S(\magenta{t})\,,
    \label{eq:neumann}\\
  \dot{L}(\magenta{t}) = -\frac{\partial_y^2 u\big|_{y=0}}{S(\magenta{t})}\,,
    \label{eq:stefan}
\end{gather}
where $\dot{L} \equiv \magenta{\dd L/\dd t}$ is the boundary
speed and $S(\magenta{t}) = (\Lambda/D) \int_0^\infty u^k\,\dd y$.
For $k = 3$ the ratio $\Lambda/D = \alpha$ is the single
dimensionless control parameter; for $k \neq 3$ one can rescale~$y$
to set $\Lambda/D$ to unity.
The Neumann condition~\eqref{eq:neumann} follows from differentiating
the mass constraint $\int u\,\dd y = 1$ in time and using
Eq.~\eqref{eq:pde_comov}; the Stefan condition~\eqref{eq:stefan} then
determines the boundary velocity $\dot{L}$.

The co-moving coordinate is discretized on a uniform grid of $N_y + 1$ points,
$y_j = j\,\Delta y$, $j = 0,1,\ldots,N_y$, with central differences
for diffusion and advection.  At the left boundary ($j = 0$) a ghost
node $u_{-1}$ enforces the Neumann
condition~\eqref{eq:neumann}: $u_{-1} = u_1 - 2\,\Delta y\,S$,
which is substituted into the stencil at $j=0$ to close the
tridiagonal system.  The curvature needed for the Stefan
condition~\eqref{eq:stefan} is then
\begin{equation}\label{eq:uyy0}
  \partial_y^2 u\big|_{y=0}
  = \frac{u_{-1} - 2u_0 + u_1}{\Delta y^2}
  = \frac{2(u_1 - u_0 - \Delta y\,S)}{\Delta y^2}\,.
\end{equation}

Time integration uses a linearized-implicit Crank--Nicolson (CN)
scheme~\cite{CrankNicolson1947}: diffusion and advection are treated with the standard CN
average (half implicit, half explicit), and the nonlinear reaction
term $(\Lambda/D)\,u^k$ is linearized as
$(\Lambda/D)\,(u^n)^{k-1}\,u^{n+1}$, where $u^n$ is the solution at
the old time level.  This yields a tridiagonal system at each step,
solved via \texttt{scipy.linalg.solve\_banded}~\cite{SciPy2020}.  After each solve,
$u_0$ is pinned to zero to enforce~\eqref{eq:abs_bc} and any negative
values are clipped.  When the cell Peclet number
$\mathrm{Pe} = |\dot{L}|\,\Delta y$ exceeds~2, the CN step is
subdivided into $\lceil\mathrm{Pe}/2\rceil$ substeps (with $\dot{L}$
and $S$ frozen) to prevent centered-advection instability.  As an
additional safeguard, $|\dot{L}|$ is capped at $\Delta y/\Delta t$
(one grid cell per step) to prevent runaway when $S\to 0$.

Mass conservation is \emph{not} explicitly enforced; the Neumann
condition~\eqref{eq:neumann} provides it approximately.  The integral
$\int u\,\dd y$ is evaluated at every step and serves to monitor the accuracy.
The pinning of $u_0 = 0$ and clipping of negative values introduce a
small mass drift; typical values are $\lesssim 1\%$ over the full
simulation time.

\section{Monte Carlo simulations}\label{sec:mc}


The lattice model is simulated with the Gillespie direct
method~\cite{Gillespie1977}.  Each step proceeds as follows:
\begin{enumerate}
\item Compute the total event rate
$W_{\rm tot} = W_{\rm hop} + W_{\rm rxn}$,
where $W_{\rm hop} = 2D_0 N$ is the total hopping rate of all $N$
particles ($D_0$~is the per-direction hopping rate) and
$W_{\rm rxn} = \sum_j w_j$ with the per-site reaction rate ($n_j$ is the number of particles at site~$j$)
\begin{equation}\label{eq:propensity}
  w_j = \frac{\Lambda}{N^{k-1}}\,(n_j)_k\,,
  \qquad
  (n)_k \equiv n(n\!-\!1)\cdots(n\!-\!k\!+\!1)\,.
\end{equation}
Here $(n)_k$ is the falling factorial and
$\Lambda = \lambda/k!$ is the rate constant defined above Eq.~(3) of the
main text.

\item Draw the waiting time $\delta t = -\ln U_1 / W_{\rm tot}$,
where $U_1$ is a uniform random number on $(0,1)$, and advance the clock.

\item With probability $W_{\rm hop}/W_{\rm tot}$, execute a
\emph{hop}: select a particle uniformly at random and move it to a
randomly chosen nearest neighbor.

\item Otherwise, execute a \emph{reaction-and-removal}: select the
reaction site~$j$ with probability $w_j/W_{\rm rxn}$, create a new
particle at~$j$, and simultaneously remove one particle from the
leftmost occupied site~$j_L(t)$.  When $n_{j_L}$ drops to zero,
$j_L$ advances to the next occupied site.  This coupled step conserves
the total particle number~$N$ exactly.
\end{enumerate}

Initial conditions are Gamma-distributed profiles across lattice
sites.  For narrow initial conditions, where the standard deviation
$\sigma_{\rm lattice}$ of the initial profile (in lattice sites) satisfies
$\sigma_{\rm lattice} \ll N$, a deterministic discretization of the
Gamma density is used; for broad initial conditions, particles are sampled stochastically to avoid rounding artifacts at low per-site occupancy.

The simulator is written in C++ (called from Python), with constant-time particle selection for hops and logarithmic-time reaction-site selection.  

\section{$k=3$: Diffusive spreading in the subcritical case $\alpha<\alpha_c$}\label{sec:subcritical}
Let us go over to the reference frame moving with the absorbing boundary, $y=x-L(t)$. The governing PDE becomes 
\begin{equation}\label{pde3movingSM}
\partial_t u =\Lambda u^3+D \partial_y^2 u +\dot{L}\partial_y u, \quad y>0\,.
\end{equation}
At $\alpha<\alpha_c$ the population spreads, and the long-time dynamics is self-similar and exhibits diffusive scaling: 
\begin{equation}
    u(y,t)=\frac{1}{L(t)}V\!\left(\frac{y}{L(t)}\right),  \label{eq:ansatzSM}
\end{equation}
where $L(t) = \beta \sqrt{Dt}$, and $\beta$ is an a priori unknown dimensionless parameter.  Plugging the ansatz (\ref{eq:ansatzSM})  into Eq.~(\ref{pde3movingSM}), we arrive at the nonlinear ODE
\begin{equation}
    V''+\alpha V^3+\gamma\left[(z+1)V'+V\right]=0,
    \qquad z>0,
    \label{eq:V_odeSM}
\end{equation}
where $z=y/L(t)$ and $\alpha=\Lambda/D$. The mass conservation reads $\int_0^{\infty} V(z) dz =1$, and the boundary condition at the absorbing wall is $V(0)=0$.  The parameter $\gamma=\beta^2/2$ plays the role of a nonlinear eigenvalue.    

Our method of numerical solution of this problem exploits the fact that $V(z)$ must decay faster than $1/z$  at $z\to \infty$, so that the normalization condition $\int_0^\infty V(z)\,dz=1$ is obeyed.  Solving the linearized version of  Eq.~(\ref{eq:V_odeSM}) , with the term $V^3(z)$ neglected,  we see that the allowed solution decays  as $\sim \exp[-(\gamma/2) (1+z)^2]$.  Imposing this large-$z$ asymptotic and its $z$-derivative on the sufficiently far numerical boundary $z=L_{\text{num}}$ and fixing $\gamma$, we can solve Eq.~(\ref{eq:V_odeSM}) in the region $z<L_{\text{num}}$, identify the first zero $z_*$ of $V(z)$ from the right, and shift the $z$-axis accordingly: $z\to z-z_*$ so as to obey the absorbing boundary condition $V(z=0)=0$. The eigenvalue $\gamma$ is selected iteratively to satisfy the normalization condition. 

\begin{figure}[t]
    \centering
    \includegraphics[width=0.85\textwidth]{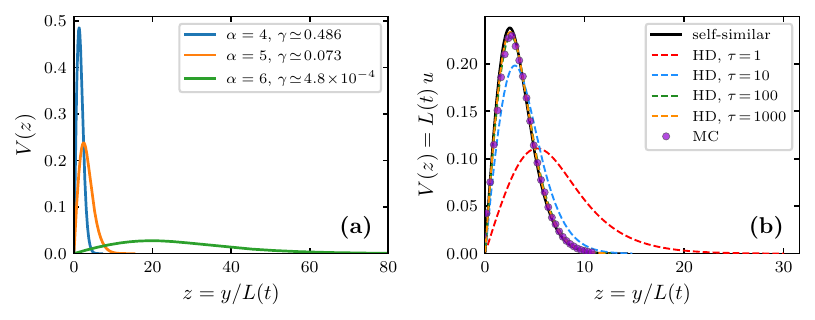}
    \caption{(a)~Numerically computed self-similar profiles $V(z)$, where $z=y/L(t)$, for the subcritical ternary reproduction with $\alpha=4$ (blue), $5$ (orange) and $6$ (green). The corresponding nonlinear eigenvalues are $\gamma_4\simeq 0.486$, $\gamma_5\simeq 0.073$, and $\gamma_6\simeq 4.8 \times 10^{-4}$.  (b)~Convergence to the self-similar profile for $\alpha=5$: the HD solution of Eq.~(\ref{pde3movingSM}), rescaled to self-similar variables $z=y/L(\magenta{t})$ and $V=L(\magenta{t})\,u$, is shown at \magenta{times $t=1$, $10$, $100$ and $1000$} (dashed lines). The black solid curve is the eigenfunction $V(z)$ of Eq.~(\ref{eq:V_odeSM}). Purple circles: MC simulation of the lattice model with $N=500$ particles at $\magenta{t}\simeq 1.3\times 10^4$.}
    \label{fig:alpha456_profile}
\end{figure}

Figure \ref{fig:alpha456_profile}(a) shows numerically computed scaling functions $V(z)$ for three values of $\alpha$:  $\alpha=4$, $5$  and $6$.  To remind the reader, $\alpha_c \simeq 7.089$, see the main text. 
The corresponding nonlinear eigenvalues are $\gamma_4\simeq 0.486$, $\gamma_5\simeq 0.073$, and $\gamma_6\simeq 4.8 \times 10^{-4}$. As one can see, $\gamma$ -- and, as a result, the coefficient $\beta =\sqrt{2\gamma}$ of the diffusive scaling $L(t) = \beta \sqrt{Dt}$ go down as $\alpha$ approaches the critical value $\alpha_c \simeq 7.089$ from below. Figure \ref{fig:alpha456_profile}(b) demonstrates convergence of the HD solution to the self-similar profile, and good agreement with MC simulations, for $\alpha=5$.

\section{$k=3$: Finite-time collapse: a road map to full solution}\label{sec:collapse}
Here we provide a road map to the construction of the full solution describing the finite-time collapse for $k=3$ and $\alpha>\alpha_c$.  We use the co-moving coordinate $y=x-L(t)$ and denote by $\Delta t = t_c - t$ the remaining time until the collapse singularity at $t=t_c$; the constants $a(\alpha)$ and $C(\alpha)$ are the nonlinear eigenvalue and the tail coefficient of the self-similar solution, respectively [see Eqs.~(13)--(15) of the main text].  The non-self-similar outer solution is described by Eq.~(12) 
of the main text with the nonlinear term neglected:
\begin{equation}\label{pdequat3b}
\partial_t u = D \partial_y^2 u +\dot{L}\partial_y u, \quad y>0.
\end{equation}
In terms of $u(y,t)$, the large-$z$ tail (15), where $z = y/\sqrt{D\Delta t}$ is the similarity variable, 
of the self-similar solution can be rewritten as
\begin{equation}
\label{statictail2}
u(y,t)\simeq\frac{C(\alpha)}{y-2a \sqrt{D\Delta t}}\,.
\end{equation}
Notice that, back in the original coordinate $x$, this near tail is, in the  leading order, \emph{static}:
$$
u(x)\simeq\frac{C(\alpha)}{x-L_c}\,,
$$
where $L_c = L(t_c)$ is the boundary position at the collapse time.
The near tail (\ref{statictail2}) needs to be matched to the outer solution in their joint validity region. As one can check directly, the near tail obeys the reduced equation
\begin{equation}
\label{kinematic}
\partial_t u =\dot{L}\partial_y u, \quad y>0,
\end{equation}
which coincides with Eq.~(\ref{pdequat3b}) with the diffusion term neglected. This is the joint region where the two solutions are both valid and can be matched.

\section{Discrete mean-field theory}\label{sec:mf}

\subsection{Lattice equations and dimensional analysis}

Consider $N$ particles on a one-dimensional lattice with spacing
$h = 1$.  Let $n_j$ denote the number of particles at site~$j$.  Each particle hops to each neighbor at rate $D_0$, and the
$kA\!\to\!(k\!+\!1)A$ reaction fires
with rate $(\Lambda/N^{k-1})\,(n_j)_k$ at site $j$ [cf.\
Eq.~\eqref{eq:propensity}].  After each birth, the leftmost particle
is removed.  In the mean-field approximation (replacing $(n_j)_k$ by
$n_j^k$), the rate equations for $k = 3$ read
\begin{align}
  \frac{\dd n_j}{\dd t}
  &= D_0\bigl(n_{j-1} + n_{j+1} - 2n_j\bigr)
    + \frac{\Lambda}{N^2}\,n_j^3\,,
    \quad j > L(t)\,,\label{eq:mf_interior}\\
  \frac{\dd n_L}{\dd t}
  &= D_0\bigl(n_{L+1} - n_L\bigr)
    + \frac{\Lambda}{N^2}\,n_L^3 - \Sigma\,,
    \label{eq:mf_left}
\end{align}
where $L(t)$ is the leftmost occupied site index,
$\Sigma = \sum_j (\Lambda/N^2)\,n_j^3$ is the total birth rate, and
the Laplacian at $j = L$ uses a reflecting boundary condition (a
particle hopping left from $L$ is reflected back).
Summing over all sites, the Laplacian telescopes to zero and the
reaction terms cancel $\Sigma$, so the total particle number
$\sum_j n_j = N$ is conserved.

Defining $u_j = n_j/N$\magenta{, setting $D = D_0 = 1$ (for $h = 1$),} and dividing
Eqs.~\eqref{eq:mf_interior}--\eqref{eq:mf_left} by $D_0$, the
equations become
\begin{align}
  \frac{\dd u_j}{\dd \magenta{t}}
  &= (u_{j-1} + u_{j+1} - 2u_j)
    + \alpha\,u_j^3\,,
    \quad j > L\,,\label{eq:mf_dimless}\\
  \frac{\dd u_L}{\dd \magenta{t}}
  &= (u_{L+1} - u_L) + \alpha\,u_L^3 - \alpha\,\Sigma_u\,,
    \label{eq:mf_dimless_L}
\end{align}
where $\Sigma_u = \sum_j u_j^3$ and $\alpha = \Lambda/D$ is the
single dimensionless parameter, matching the main text.  The mass
constraint becomes $\sum_j u_j = 1$.
Since only $\alpha$ appears in the dimensionless
equations~\eqref{eq:mf_dimless}--\eqref{eq:mf_dimless_L}, the
lattice wave velocity can only depend on $\alpha$:
\begin{equation}\label{eq:vscale}
  v = D_0\,\nu(\alpha)\,,
  \qquad\text{equivalently}\quad
  \frac{v}{D} = \nu(\alpha)\,,
\end{equation}
where $\nu(\alpha)$ is a universal dimensionless function.
The quantity $c/(D_0 h) = \nu$ is plotted in Fig.~3(f) of the main
text.

Between site-advance events, the system of Eqs.~\eqref{eq:mf_dimless}--\eqref{eq:mf_dimless_L} is a smooth ODE on a fixed set of sites.  When $n_{L(t)}$ reaches zero, $L$ advances to $L + 1$.  At this instant, all occupancies are continuous -- the only discontinuity is in the time derivatives.  Specifically, $\dd u_{L+1}/\dd\magenta{t}$ jumps: before the advance, site $L\!+\!1$ has the interior Laplacian $u_L + u_{L+2} - 2u_{L+1} = u_{L+2} - 2u_{L+1}$ (since $u_L = 0$); after the advance, it inherits the reflecting Laplacian $u_{L+2} - u_{L+1}$ and the removal term $-\alpha\,\Sigma_u$. The jump in $\dd u_{L+1}/\dd\magenta{t}$ is $u_{L+1} - \alpha\,\Sigma_u$.
The system is therefore a well-posed piecewise-smooth initial-value problem, solved by concatenation: integrate the smooth ODE until $u_L = 0$ (detected by an event function), then switch to the new system with $L \to L + 1$ and continue.  The occupancies are continuous across each switch.

\begin{figure}[t]
\includegraphics[width=\columnwidth,clip=]{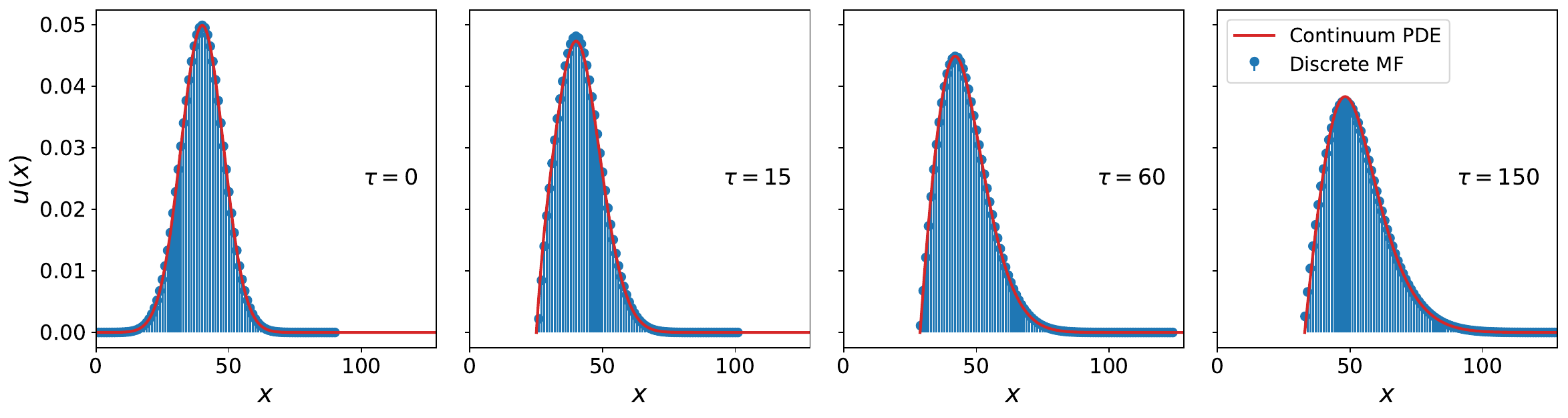}
\caption{\label{fig:mf_vs_pde}%
  Discrete mean-field profiles (blue stems) vs.\ continuum HD solution
  (red lines) at $\alpha = 5$ for four times
  $\magenta{t} = 0, 15, 60, 150$.  Both start from a Gaussian with
  $\sigma = 8$ and center $x_0 = 40$.  Mass annotations show that the
  discrete MF conserves mass exactly ($\textstyle\sum_{j=0}^M u_j = 1.0000$) while the HD solver
  maintains $\textstyle\sum_{j=0}^M u_j \lesssim 1.02$ without explicit mass correction.}
\end{figure}

Figure~\ref{fig:mf_vs_pde} validates this discrete mean-field model
by comparing the time-dependent solution of
Eqs.~\eqref{eq:mf_dimless}--\eqref{eq:mf_dimless_L} with the
continuum HD solver of Sec.~\ref{sec:pde} at $\alpha = 5$ (the
sub-critical diffusive regime), starting from a Gaussian initial
condition with $\sigma = 8$ centered at site $x_0 = 40$.  The two
solutions are in close agreement, confirming that the discrete lattice
equations reduce to the continuum HD equations when the density profile spans
many lattice sites.

\subsection{Traveling-wave formulation}

The traveling-wave (TW) ansatz $u_j(\magenta{t}) = \phi(j - \nu\magenta{t})$, with
$\xi = j - \nu\magenta{t}$ a continuous variable, yields the advance-delay
functional differential equation
\begin{equation}\label{eq:advdelay}
  -\nu\,\phi'(\xi) = \phi(\xi\!-\!1) + \phi(\xi\!+\!1) - 2\phi(\xi)
    + \alpha\,\phi(\xi)^3
    - \delta_{\xi,0}\,\alpha\,\Sigma_\phi\,,
\end{equation}
where $\Sigma_\phi = \sum_j \phi(j)^3$ and $\nu = v/D_0$ is the
dimensionless lattice wave speed~\eqref{eq:vscale}.
The function $\phi(\xi)$ must be defined as a \emph{continuous}
function on all of $\mathbb{R}$, not merely at integer points; the
physical occupancies are $u_j(\magenta{t}) = \phi(j - \nu\magenta{t})$, i.e., the
continuous profile sampled at integer sites that slide in time.  The
derivative $\phi'(\xi)$ appearing in~\eqref{eq:advdelay} is the
spatial derivative of this smooth interpolating profile, not the
(potentially discontinuous) time derivative $\dd u_j/\dd\magenta{t}$ at a
single lattice site discussed above.

Evaluating~\eqref{eq:advdelay} at integer sites $\xi = j$ and
approximating $\phi'(j)$ by the forward difference
$\phi(j\!+\!1) - \phi(j) = u_{j+1} - u_j$ gives, after truncation
at $j = M$ with $u_{M+1} = 0$, the algebraic system
\begin{align}
  j = 0:&\quad (u_1\!-\!u_0) + \alpha\,u_0^3
    - \alpha\,\Sigma_u + \nu(u_1\!-\!u_0) = 0\,,
    \label{eq:tw0}\\
  1 \leq j < M:&\quad (u_{j-1}\!+\!u_{j+1}\!-\!2u_j) + \alpha\,u_j^3
    + \nu(u_{j+1}\!-\!u_j) = 0\,,
    \label{eq:twj}\\
  j = M:&\quad (u_{M-1}\!-\!2u_M) + \alpha\,u_M^3
    - \nu\,u_M = 0\,,
    \label{eq:twM}\\
  \text{mass:}&\quad \textstyle\sum_{j=0}^M u_j = 1\,.
    \label{eq:twmass}
\end{align}
This is a system of $M + 2$ equations in $M + 2$ unknowns
$(u_0, \ldots, u_M, \nu)$, solved numerically with
\texttt{scipy.optimize.fsolve}~\cite{SciPy2020} using an analytical Jacobian and a
random-seed scanning strategy ($\sim\!2000$ Dirichlet-distributed
initial guesses).

\section{$k=4$: Logarithmic corrections to diffusive spreading}\label{sec:k4log}

As shown in the main text, for $k=4$ the reproduction term in Eq.~(3) is asymptotically subleading relative to diffusion.  Nevertheless, it produces logarithmic corrections to the purely diffusive behavior, which we derive here.

Differentiating the mass constraint~(2) and using the boundary condition $u(L(t),t)=0$ yield the exact flux balance
\begin{equation}\label{eq:flux_balance_k4}
  \magenta{D\,\partial_x u(L(t),t)
  = \Lambda\int_{L(t)}^{\infty} u^4(x,t)\,\dd x}\,.
\end{equation}
Next, define the center of mass $M(t) = \int_{L(t)}^{\infty} x\,u(x,t)\,\dd x$.  Differentiating this relation, using Eq.~(3) with $k=4$, integrating by parts, and applying Eq.~\eqref{eq:flux_balance_k4} yield the exact identity
\begin{equation}\label{eq:Mdot_k4}
  \magenta{\dot{M}(t)
  = \Lambda\int_{L(t)}^{\infty}\bigl[x - L(t)\bigr]\,u^4(x,t)\,\dd x}\,.
\end{equation}
This identity shows that the reproduction in the bulk and the absorption at the boundary produce a
rightward drift of the center of mass.

Assuming that the diffusion dominates over the reproduction, we can approximate the long-time bulk profile of $u(x,t)$ by a unit-mass Gaussian centered at $M(t)$:
\begin{equation}\label{eq:gaussian_k4}
  \magenta{u(x,t) \simeq
  \frac{1}{\sqrt{4\pi D t}}\,
  \exp\!\left[-\frac{(x-M)^2}{4Dt}\right]}.
\end{equation}
\magenta{Now we can evaluate the integral in Eq.~(\ref{eq:flux_balance_k4}):}
\begin{equation}\label{eq:u4int_k4}
  \magenta{\int_{-\infty}^{\infty} u^4(x,t)\,\dd x
  = \frac{1}{16\pi^{3/2}(Dt)^{3/2}}}\,.
\end{equation}
where we have moved the lower limit to $-\infty$ because the boundary $x=L(t)$ is located far in the left Gaussian tail.

\begin{figure}[t]
  \centering
  \includegraphics[width=0.45\textwidth]{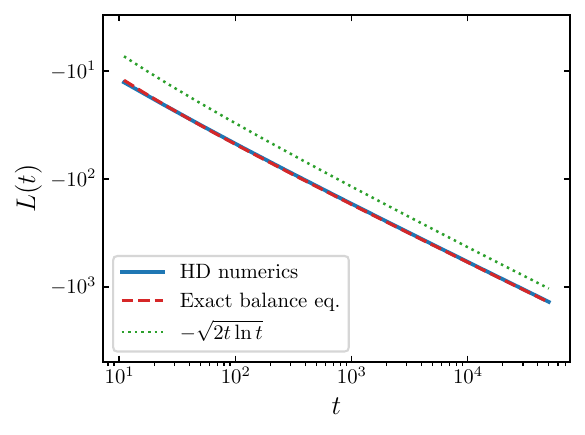}
  \caption{Boundary position $L(t)$ for quaternary reproduction ($k=4$) in units $D=\Lambda=1$, in the log scale.
  Solid line: HD numerics; dashed line: $L(t) = M(t) - A(t)$ with $A(t)$ from the numerical solution of Eq.~\eqref{eq:balance_implicit_k4} at different~$t$ and $M(t)$ from integrating Eq.~\eqref{eq:Mdot_k4}; dotted line: the leading asymptotic $-\sqrt{2t\ln t}$ from Eq.~\eqref{eq:L_k4}.}
  \label{fig:k4_logcorr}
\end{figure}

Let $A(t) = M(t) - L(t) > 0$ denote the distance between the center of mass $x=M(t)$ and the boundary $x=L(t)$.  Equation~(\ref{eq:flux_balance_k4}) yields the flux to the boundary,
$D\,\partial_x u(L,t) \simeq [A/(4t\sqrt{\pi Dt})]\exp[-A^2/(4Dt)]$.  Equating this flux to the right-hand side of \eqref{eq:flux_balance_k4} and using Eq.~(\ref{eq:u4int_k4}) gives the implicit balance equation
\begin{equation}\label{eq:balance_implicit_k4}
  \magenta{\sqrt{s}\,e^{-s} = \frac{\Lambda}{8\pi D\sqrt{Dt}}\,,
  \qquad s \equiv \frac{A^2}{4Dt}}\,,
\end{equation}
\magenta{which, to leading logarithmic accuracy, yields}
\begin{equation}\label{eq:A_k4}
  \magenta{A(t) \sim \sqrt{2Dt\ln t}}\,.
\end{equation}
\magenta{Since $u^4$ is concentrated near $x = M(t)$, Eq.~\eqref{eq:Mdot_k4} gives $\dot{M} \simeq \Lambda\,A(t)\int_{L(t)}^{\infty} u^4(x,t)\,\dd x$.  Substituting \eqref{eq:u4int_k4} and \eqref{eq:A_k4}, we obtain}
\begin{equation}\label{eq:Mdot_asymp_k4}
  \magenta{\dot{M}(t) \sim
  \frac{\Lambda\sqrt{2}}{16\pi^{3/2} D}\,
  \frac{\sqrt{\ln t}}{t}}\,.
\end{equation}
Integrating this in time using $\int t^{-1}\sqrt{\ln t}\,\dd t = \tfrac{2}{3}(\ln t)^{3/2}$ yields the leading rightward center-of-mass drift $M(t) \sim \frac{\Lambda\sqrt{2}}{24\pi^{3/2} D}\,(\ln t)^{3/2}$. Combining this with Eq.~\eqref{eq:A_k4} gives the boundary position versus time:
\begin{equation}\label{eq:L_k4}
  \magenta{L(t) = M(t) - A(t) \simeq -\sqrt{2Dt\ln t} + \frac{\Lambda\sqrt{2}}{24\pi^{3/2} D}\,(\ln t)^{3/2} + \cdots}\,.
\end{equation}
The first term in this asymptotic dominates. 
Figure~\ref{fig:k4_logcorr} compares $|L(t)|$ obtained in the full HD solution (solid line) with the leading asymptotic $\sqrt{2Dt\ln t}$ from Eq.~\eqref{eq:L_k4} (dotted line) and with the more accurate asymptotic, obtained by solving the balance equation~\eqref{eq:balance_implicit_k4} numerically for $A(t)$ at each~$t$ and using $M(t)$ determined above (dashed line).  The more accurate asymptotic captures the HD numerics 
to within $\sim\!1\%$.

\end{document}